\documentclass[journal=jpccck,manuscript=article]{achemso}
\usepackage[version=3]{mhchem} 
\usepackage{color}
\usepackage{multirow}
\usepackage{amssymb}

\author{Guilherme S. L. Fabris}
\affiliation[UFRN]
{Postgraduate Program in Materials Science and Engineering, Federal University of Pelotas, 96010-610, Pelotas, RS, Brazil}

\author{D. S. Galvão}
\affiliation[IFGW]
{Applied Physics Department, State University of Campinas, 13083-970, Campinas SP, Brazil}
\email{galvao@ifi.unicamp.br}

\author{Ricardo Paupitz}
\affiliation[UNESP-RC]
{Sao Paulo State University - UNESP, Physics Department, CEP-13506-900 Rio Claro, SP, Brazil}
\email{ricardo.paupitz@unesp.br}

\title{Reversible Actuation of $\alpha$-Borophene Nanoscrolls}

\abbreviations{IR,NMR,UV}

\begin{document}







\begin{abstract}

In this work, we proposed and investigated the structural and electronic properties of boron-based nanoscrolls (armchair and zigzag) using the DFTB+ method. We also investigated the electroactuation process (injecting and removing charges). A giant electroactuation was observed, but the results show relevant differences between the borophene and carbon nanoscrolls. The molecular dynamics simulations showed that the scrolls are thermally and structurally stable for a large range of temperatures (up to 600K) and the electroactuation process can be easily tuned and can be entirely reversible for some configurations.

\end{abstract}

\section{Introduction}

Since the experimental realization of graphene\cite{novoselov2004}, the search for low-dimensional materials has significantly increased. These materials' interest is partly due to their special properties and wide range of potential applications\cite{novoselov2009,wu2008epitaxial}. Among the low-dimension structural families, carbon nanoscrolls, which can be described as tubular structures made from a continuously wrapped graphene sheet into a spiral shape (papyrus-like), have been studied for a few years and now due to new works are of renewed  interest\cite{Coluci_scrolls_2006_electroactuation,Pugno_scrolls_2010_review}. They have been synthesized using different methods and present interesting and unique structural, mechanical, and electronic properties\cite{Sun_scrolls_2011_raman,Dong_scrolls_2010_fabrication,Pugno_scrolls_2010_core_size,Pugno_scrolls_2010_water_channels}.

The material science revolution created by the advent of graphene,  stimulated the interest in new materials of various types. In this context, many new inorganic materials, structurally related to carbon nanostructures, were proposed and investigated in the last years as, for example, the BN analogs of graphene, nanotubes\cite{Golberg_hBN_tubes_2010} and nanoscrolls\cite{Perim_scrolls_BN_2009}. The graphene-like structures of other elements have been proposed (and some already experimentally realized) like silicene, germanene, stanene, gallium nitride, among others \cite{Phosphorene-tomanek,bismuthene2016,Golberg982441,Zhang2018,Rubio945081}. Some of these materials' chemical and physical properties are appealing for technological applications, as they exhibit good electrical and thermal conductivities, stiffness, etc., which make them good candidates for potential use in molecular sieves, energy storage devices or as part of electronic devices.

\begin{figure}[h]
\includegraphics[width=5.0 in, keepaspectratio]{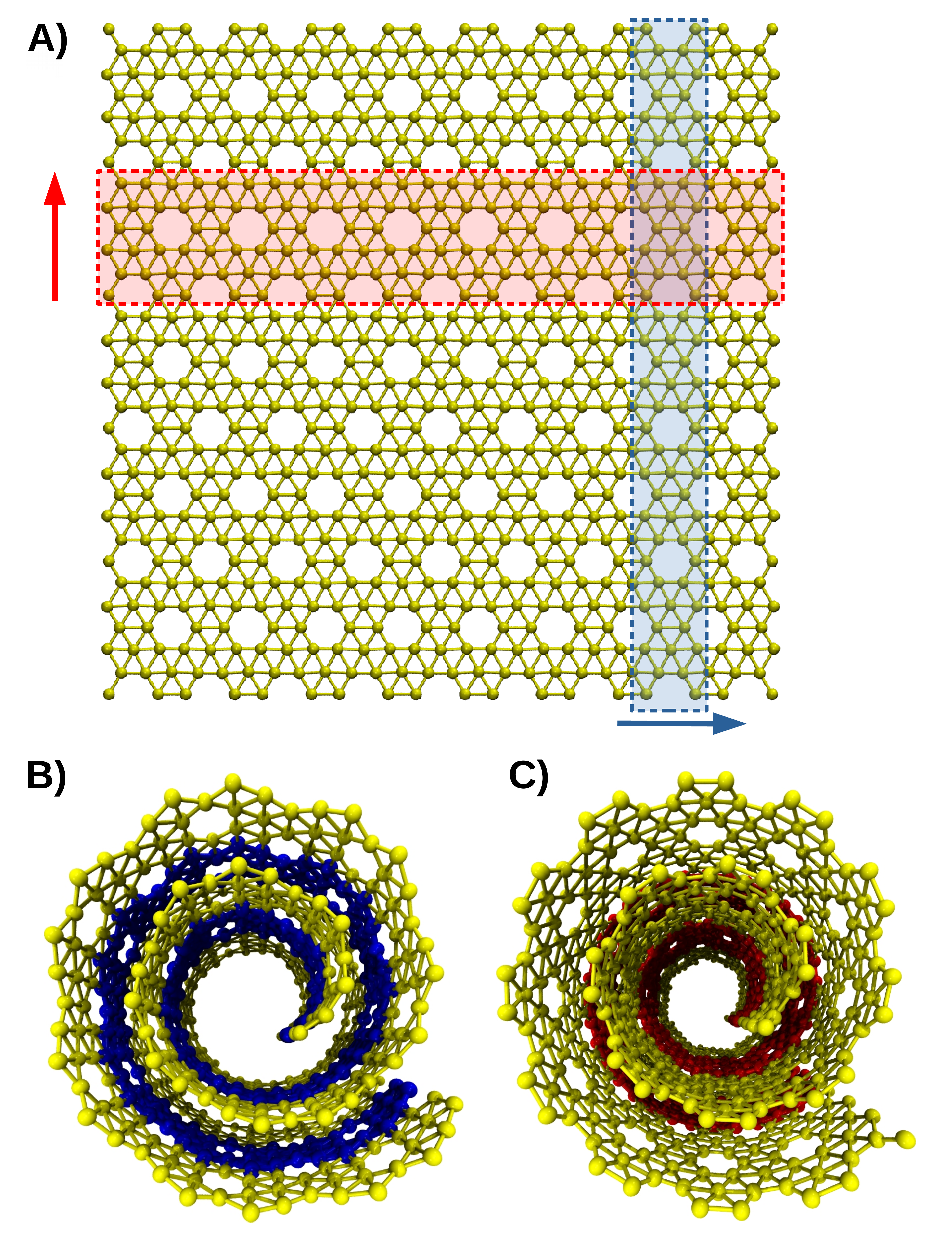}
\caption{(A) $\alpha-$Borophene monolayer with the unit cells for the armchair and zigzag nanostrips highlighted. The blue (red) arrow indicates the axis direction for the armchair (zigzag) nanoscroll after the rolling up process. (B) Internal view of an armchair nanoscroll, in which blue atoms and bonds indicate the rolled-up unit cell corresponding to the highlighted area with the same color for an armchair nanostrip in (A). (C) Internal view of a zigzag nanoscroll, where red atoms and bonds indicate the rolled-up unit cell corresponding to the highlighted area with the same color for a zigzag nanostrip in (A). }
\label{fig:cell}
\end{figure}

Boron-based materials are of significant interest among these structures, mainly because this element can form many different allotropic structures \cite{borophene-yakobson2018,boron-cluster,strain-borophene}. These features pose this element as a natural candidate to be investigated as building blocks for new low-dimensional materials \cite{borophene-dirac-fermions,borophene-review-2019,strain-borophene-2016}.  Of particular interest are the borophenes, structures composed of monolayers of boron atoms (see Figure (\ref{fig:cell}-A)). They are structurally stable \cite{borophene_theoretical_2012,Yakobson_2015} and several 2D configurations with very close cohesive energies were identified, despite their clear structural differences\cite{Yakobson_2012}. The experimental realization of some borophenes was recently reported \cite{borophene_2105,borophene_2016}.

Concerning carbon nanoscrolls, it is also worth mentioning that Rurali et al. \cite{Coluci_scrolls_2006_electroactuation} reported giant charge actuation.  After charge injection, the carbon nanoscrolls exhibit a significant change of their axial diameter, revealing a form of electroactuation one order of magnitude larger than that found for carbon nanotubes under similar conditions. 


A natural question is whether borophene nanoscrolls can be formed, if they are structurally stable, and if they can exhibit giant electroactuation. In this work, we have investigated the structural, electronic, topological, and electroactuation properties of $\alpha$-borophene nanoscrolls (Figure (\ref{fig:cell})) using the Tight Binding Density Functional Theory (DFTB) method \cite{manzano2012,Elstner1998,Kubar2013}. The nanoscroll structures were obtained by rolling up $\alpha$-borophene sheets along two different directions, namely zigzag and armchair (see Figure (\ref{fig:cell}-A)). Our results show that $\alpha$-borophene nanoscrolls are structurally stable and exhibit a significant (and reversible, upon removing the injected charges) electroactuation.

\section{Methods}

Computational simulations were carried out within a DFTB approximation in its self-consistent-charge (SCC)-DFTB version\cite{Elstner1998}, as implemented in the DFTB+ code\cite{dftb2020}. Using this approximation, we can carry out electronic structure calculations, geometrical optimizations, and molecular dynamics simulations with a relatively low-computational cost (in comparison to full DFT methods). DFTB+ \cite{manzano2012,Elstner1998,Kubar2013} allows quantum simulations of the electronic and structural properties for relatively large systems with a computational cost similar to other tight-binding methods but with a precision that can be comparable to pure DFT methods in some cases~\cite{manzano2012,dftb_performance}. The DFTB+ is based on a second-order expansion of the Kohn-Sham energy, as described by several authors~\cite{koskinen2009beginners,dftb2020}. In the present work, we adopted the {\it Borg} parametrization\cite{GrundktterStock2012}, developed for molecular systems, which included boranes and pure boron nanostructures. 

DFTB+ Molecular Dynamics (MD) simulations were carried out for several temperature values. We used the Verlet algorithm and a Nose-Hoover thermostat~\cite{nose-hoover}. Atomic structure visualizations and the MD movies (see the Supplementary Materials) were obtained using the Tachyon ray tracing library, built into VMD software~\cite{vmd,tachyon}, and the VESTA software\cite{vesta} was used to obtain orbital visualizations.


For the geometry optimization processes, we used a conjugate gradient algorithm as implemented in DFTB+ code with convergence criteria for the SCC interactions and force differences smaller than $10^{-8}$ and $10^{-5}$ a.u., respectively. The first Brillouin zone was sampled with a 1x6x1 Monkhorst-Pack grid\cite{monkhorst_pack}, in which the $6$ folding is along the periodic direction of the structure. The isosurfaces were modeled using an isovalue resolution of 0.005 $\AA^{-1}$.

The $\alpha$-borophene nanoscrolls (Figure 1) were generated using $\alpha$-borophene square monolayers with a length $L\approx 98\AA$, considering zigzag or armchair edges depending on the case. Full geometry optimizations were carried out for these monolayers before rolling them up (using a homemade computational code) to generate the initial nanoscroll configurations: interlayer distance of $4\AA$ and a curvature radius of $10\AA$. The thus generated nanoscrolls were subsequently fully optimized.

\section{Results and discussion}

As mentioned above, the nanoscroll structures were generated from rolling up the  2D $\alpha$-borophene rectangular supercells, which were subsequently geometry optimized using DFTB+. For the 2D $\alpha$-borophene, our results indicate a P$_{6}/$mmm space group, with eight atoms in the unit cell, lattice parameters $a=b=5.19\AA$, and B-B bond lengths ranging from  $1.75$ to $1.84\AA$. These results are in good agreement with those reported in the literature\cite{peng_2016}. 

Two types of supercells were considered for the rolling-up process, as shown in Figure (\ref{fig:cell}-A). As indicated, there were two directions of rolling up, which we call {\it armchair} (Figure \ref{fig:cell}-B) and {\it zigzag}(Figure \ref{fig:cell}-C) directions, respectively.
In the examples shown in Figure \ref{fig:cell}, armchair and zigzag terminations were used to make 2-turns nanoscrolls, which have a starting interlayer distance of $\sim 4\AA$ along with an initial radius of $r\sim 10\AA$. For the non-periodic directions, for the rectangular cell lengths, we considered $97.3 \times 8.9\AA$ for the armchair and $97.4 \times 5.2\AA$ for the zigzag nanoscrolls, respectively.

After the full geometry optimization processes, the obtained interlayer distances for the zigzag and armchair nanoscrolls were $d=4.01\AA$ and $d=3.97\AA$, respectively. Both present small distortions over the structure, in particular at the borders. In the innermost layer, small distortions towards the neighboring layer were observed for the zigzag nanoscrolls, with the formation of interlayer chemical bonds due to dangling bonds at the borders of the generated 2D structures. The armchair and zigzag nanoscrolls  minimum(maximum) diameters after the optimization were $\sim 18 (22)$ and $\sim 19 (21)\AA$, respectively. Both structures underwent a slight flattening, differently from what was observed for carbon nanoscrolls, which seem to have maintained their circular shape and a diameter of $\sim 20\AA$.\cite{Rurali_2006}

To help to determine the borophene scroll structural stability, we calculated the relative formation energies in relation to 2D $\alpha-$Borophene. The results are presented in Table \ref{table:energies}. We can see that neutral nanoscrolls have slightly lower energy than their corresponding 2D configurations (nanostrips), with the armchair configuration having a slightly lower energy. The scrolled armchair configuration also presents a higher energy gain in relation to the zigzag one (0.061 and 0.045 eV/atom, respectively). However, none of these structures has a lower energy than infinite borophene. Similarly to the case of carbon nanoscrolls \cite{Braga2004}, where the scrolled configuration has lower energy than the planar configuration, the same occurs with the borophene scrolls. This implies that the energy gain from the van der Waals interactions (due to the overlapping) overcomes the bending energy cost.

\begin{table}[]
\caption{Charge injection (q), the distance between inner and outer borders ($d_{tips}$), total energy ($E_{tot}$) and relative energy ($E_{rel}$) of armchair and zigzag nanoscrolls. As reference values, pristine borophene and both nanostrip conformation values are also shown. Rows marked with $*$ indicate the onset of a detachment of the external border.} 
\begin{tabular}{cllll}
\hline
                    & $ q(e/cell)/n(e/atom) $ & $d_{tips}(\AA)$ & $E_{tot} (eV/atom)$ & $E_{rel} (eV/atom)$        \\ \hline
\multirow{5}{*}{AC} & 0/0.0    &    7.48    & -28.953 &  -0.061 \\
                    & -1/0.0056   &    7.78    & -29.025
 &  -0.134
 \\
                    & -2/0.0112   &    8.64    & -29.064
 &  -0.172
 \\
                    & -3/0.0169   &    7.97    & -29.078
 & -0.186
 \\
                    & -4/0.0225   &    *       & * &           *   \\
                    \hline
\multirow{5}{*}{ZZ} & 0/0.0    &    8.13    & -28.937
 &  -0.033
 \\
                    & -1/0.0033   &    7.66    & -29.014
 &  -0.110
 \\
                    & -2/0.0066   &    8.09    & -29.028
 &  -0.124
 \\
                    & -3/0.0099   &    8.30    & -29.043
 &  -0.139
 \\
                    & -4/0.0133   &    8.27    & -29.056
 &  -0.152
 \\
                    & -6/0.0199   &    8.40    & -29.080
 &  -0.176
 \\
                    & -7/0.0232   &    *    & * &  * \\
                    \hline
Borophene           &  -   &    -       & -29.276
 &         -         \\
Nanostrip AC           &  -   &    -       & -28.892
 &         -         \\
Nanostrip ZZ        &  -   &    -       & -28.904
 &         -         \\
                    \hline
\end{tabular}
\label{table:energies}
\end{table}

\begin{figure}[h]
\includegraphics[width=6.0 in, keepaspectratio]{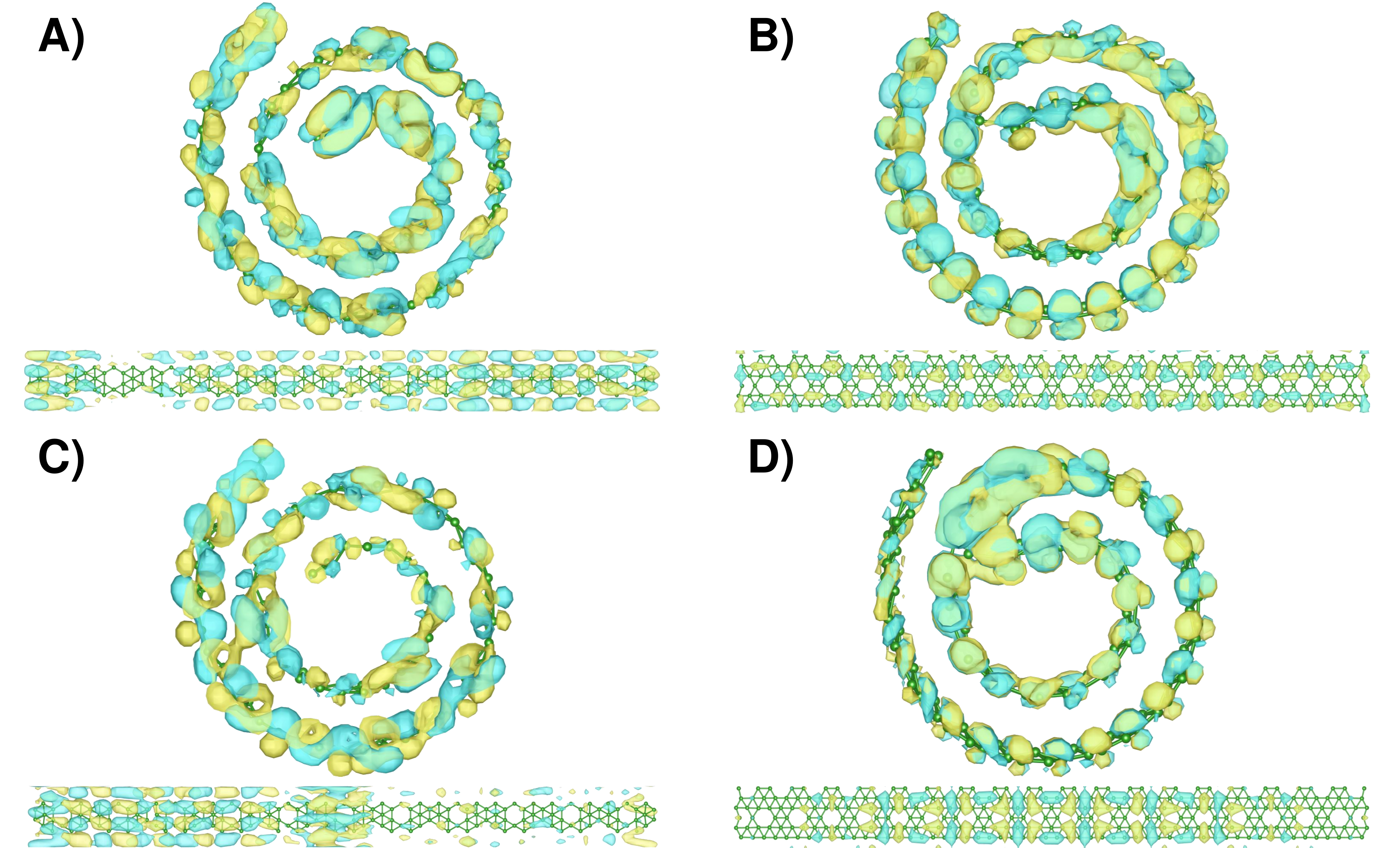}
\caption{(A) and (B) the Highest Occupied Crystalline Orbital (HOCO) of the armchair and zigzag nanoscrolls and nanostrip, respectively. (C) and (D), the corresponding Lowest Unoccupied Crystalline Orbitals (LUCO). For reference, the corresponding HOCO/LUCO of the planar configurations are also shown.}
\label{fig:AC_ZZ_HOMO_LUCO}
\end{figure}

In order to better understand the electronic changes induced by the scrolling processes, we can compare the frontier orbitals, namely HOCO (Highest Occupied Crystal Orbital) and LUCO (Lowest Unoccupied Crystal Orbital) for the planar and scrolled configurations. The results are presented in Figure \ref{fig:AC_ZZ_HOMO_LUCO}.

For the armchair-type case, the rolling process changes the HOCO patterns mainly at the borders, exhibiting a higher 
density in those regions, while the LUCO shows an increase in the orbital density from the inner to the outer layers. 
The zigzag structure presents a HOCO orbital that is more homogeneous, except for the first half turn. The LUCO has a higher density in the inner region, with a distribution extending over the different layers. 

\begin{figure}[h]
\includegraphics[width=6.5 in, keepaspectratio]{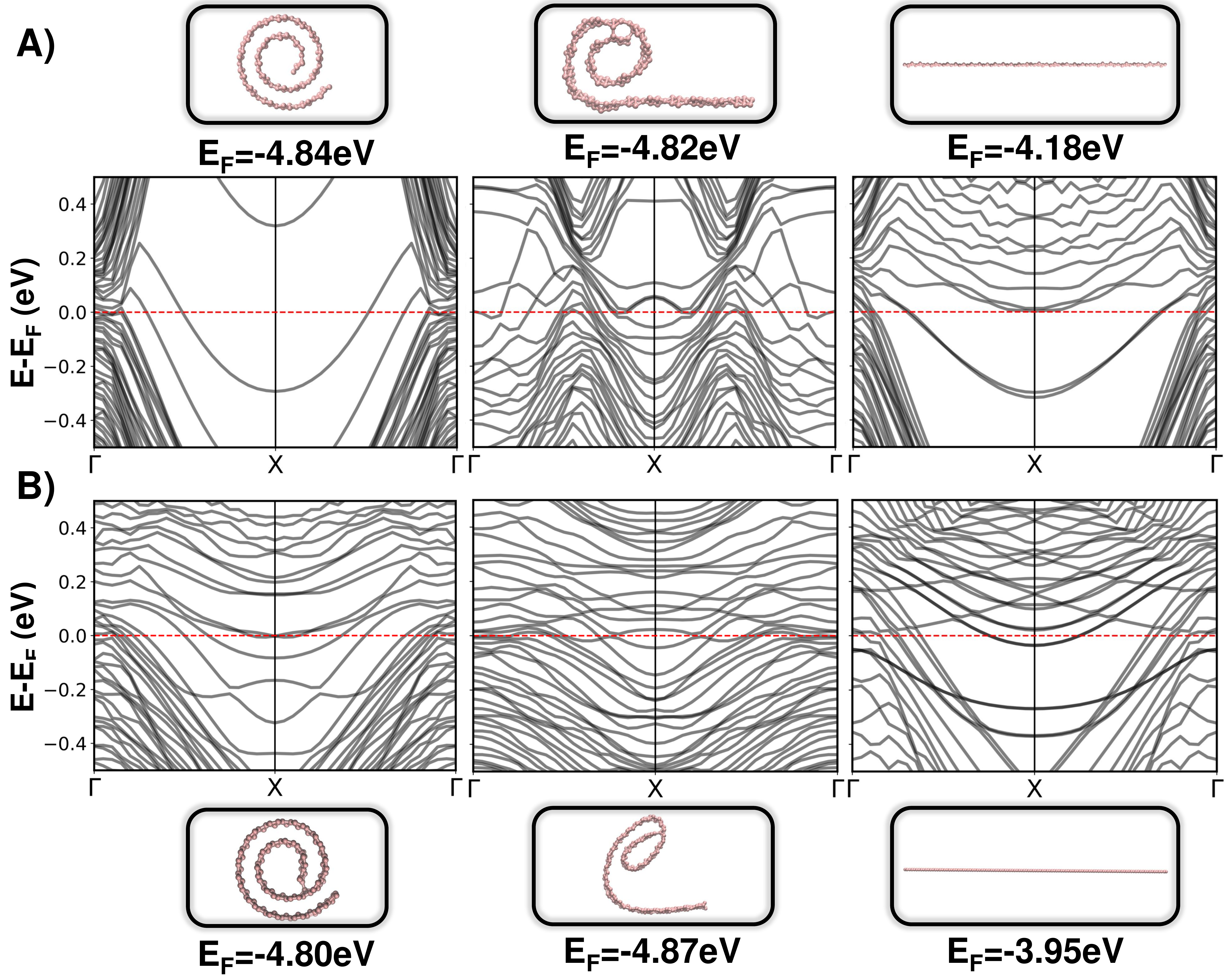}
\caption{Electronic band structure of the borophene nanoscroll, partially unrolled nanoscroll, and the corresponding armchair (A) and zigzag (B) nanostrips.}
\label{fig:scroll_BS}
\end{figure}

As can be seen from Figure \ref{fig:AC_ZZ_HOMO_LUCO}, the rolling process significantly changes the frontier orbitals in relation to the planar configurations, but as can be inferred from the electronic band structure (see Figure \ref{fig:scroll_BS}), some electronic features, such as the metallic behavior observed for the 2D structures, are preserved.

For both armchair and zigzag nanoscrolls, the number of energy bands around the Fermi level is decreased, and there is a contraction of inner bands caused by the surface bending and interaction among layers.

Considering the fact that carbon nanoscrolls present significant electroactuation \cite{Coluci_scrolls_2006_electroactuation}, it is natural to investigate whether borophene nanoscrolls can present similar behavior. For this analysis, we carried out further geometrical optimizations of the borophene nanoscrolls, injecting a different number of electrons into the system.


In Table \ref{table:energies}, we present the results of relative energies for neutral and charged nanoscrolls in both chiralities, namely zigzag and armchair. We can see from Table \ref{table:energies} that there are only slight changes in relative energy values with the number of injected electrons per unit cell up to a threshold, namely $q=4$ electrons/cell for AC and $q=7$ electrons/cell for ZZ. These threshold values indicate the maximum charge injection without significant changes in geometrical configurations.

\begin{figure}[h]
\includegraphics[width=5.0 in, keepaspectratio]{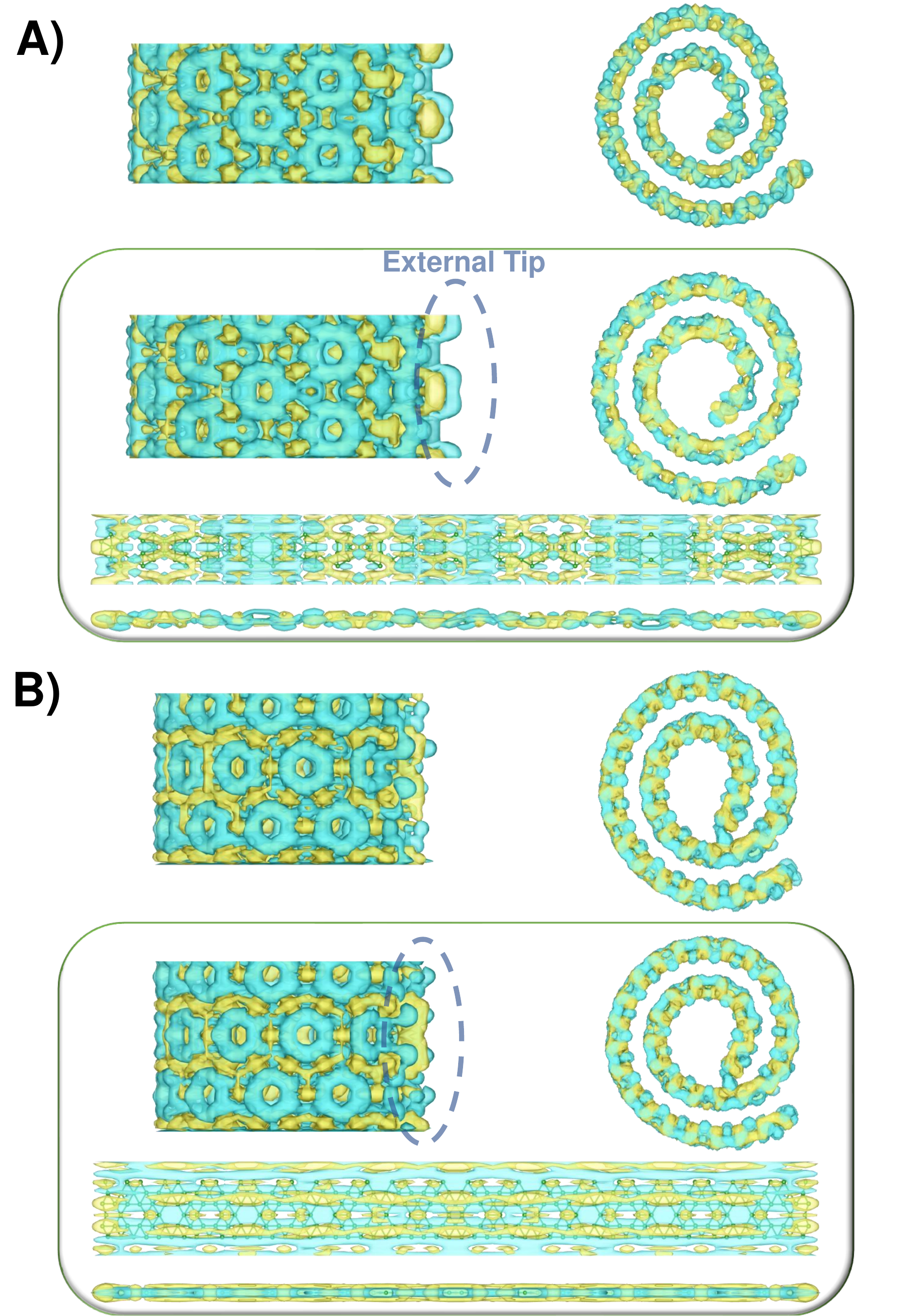}
\caption{Charged (top) and neutral (bottom) $\alpha-$Borophene nanoscrolls. (A) armchair and (B) zigzag conformation. For reference, the corresponding planar nanostrip configurations are also shown.} 
\label{fig:chargediff}
\end{figure}

In Figure \ref{fig:chargediff}, representative results for charged and neutral nanoscrolls are shown. Green and yellow indicate the regions with high and low charge density values respectively.  In such representations, one can see that charge injection modifies the charge distributions on various regions of the rolled surfaces.
For the armchair system, charge injection can modify the outermost tip of the nanoscroll, and the layers experience a repulsive force, resulting in larger bond length ($d_{BB}>1.9 \AA$) and layer separation values (see Table \ref{table:energies}). For charging of $n\ge 0.0225$e/atom, the nanoscroll starts to unroll. Similar behavior was observed in the case of zigzag chirality with a threshold $n=0.0232$e/atom, and the complete unrolling process was prevented by the formation of a few interlayer bonds. An incomplete unrolling process, in some cases, can be associated with a stabilizing effect caused by a geometrical alignment between hexagons in subsequent layers. Such alignment allows the formation of B-B chemical bonds, especially in the presence of thermal kinetic energy, which will be discussed below in the MD simulations. It was also verified that if, instead of injecting, we remove charges from the neutral nanoscroll, the behavior is not the same, and the nanoscroll just expands and then stabilizes, which may be linked to the $2p_{z}$ empty orbitals.

Comparatively, the behavior found for the borophene nanoscrolls is quite similar to that described for the carbon nanoscrolls, in which the charge injection was studied by Braga et al.\cite{Braga2004} and Rurali et al.\cite{Rurali_2006}, who found that a $\sim 5.5\%$ of charge injection changes the diameter in $ \sim 2.5\%$. However, the electroactuation response of the borophene nanoscrolls seems to be higher, as it starts to unroll at $2.2\%$ charge injection, and it is easily controlled to re-roll as found in the MD simulations.   

\begin{figure}[h]
\includegraphics[width=7.0 in, keepaspectratio]{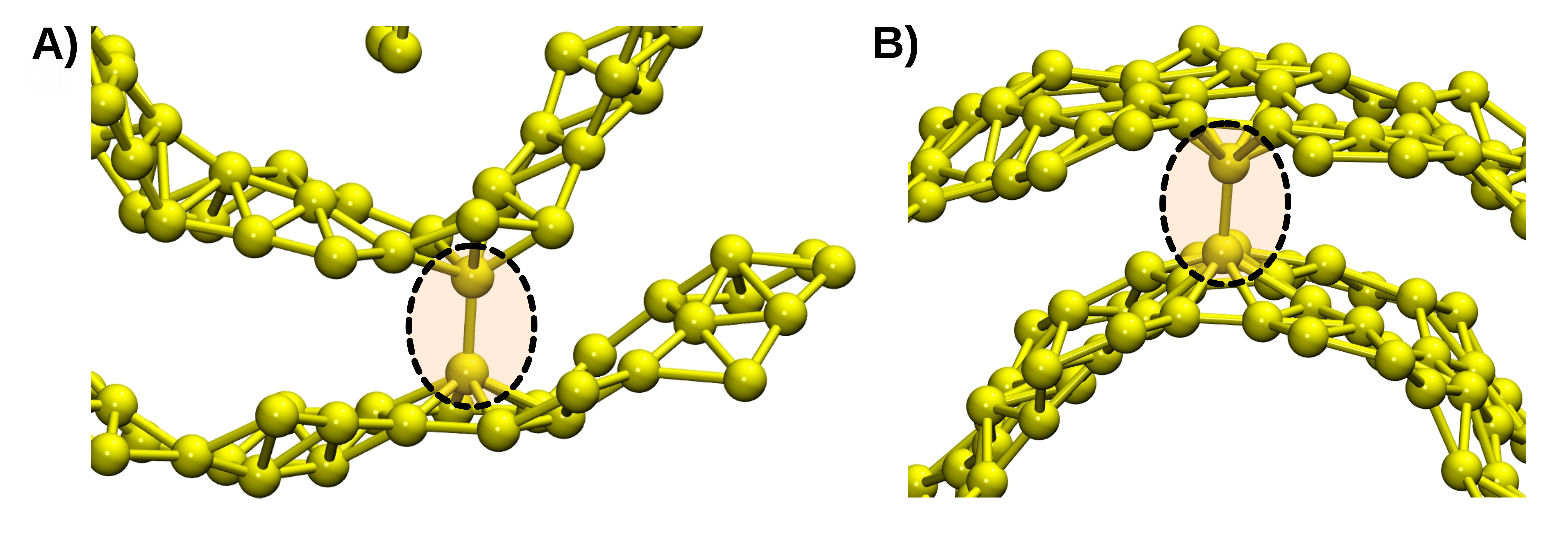}
\caption{(A) Interlayer bond connecting atoms from the hexagonal rings in an armchair nanoscroll observed during MD simulations with no charges added to the system. 
(B) Interlayer bond connecting atoms from the hexagonal rings observed during the MD simulations of a zigzag nanoscroll with no charges added. }
\label{fig:interlayer-bonds}
\end{figure}

\begin{figure}[h]
\includegraphics[width=5.0 in, keepaspectratio]{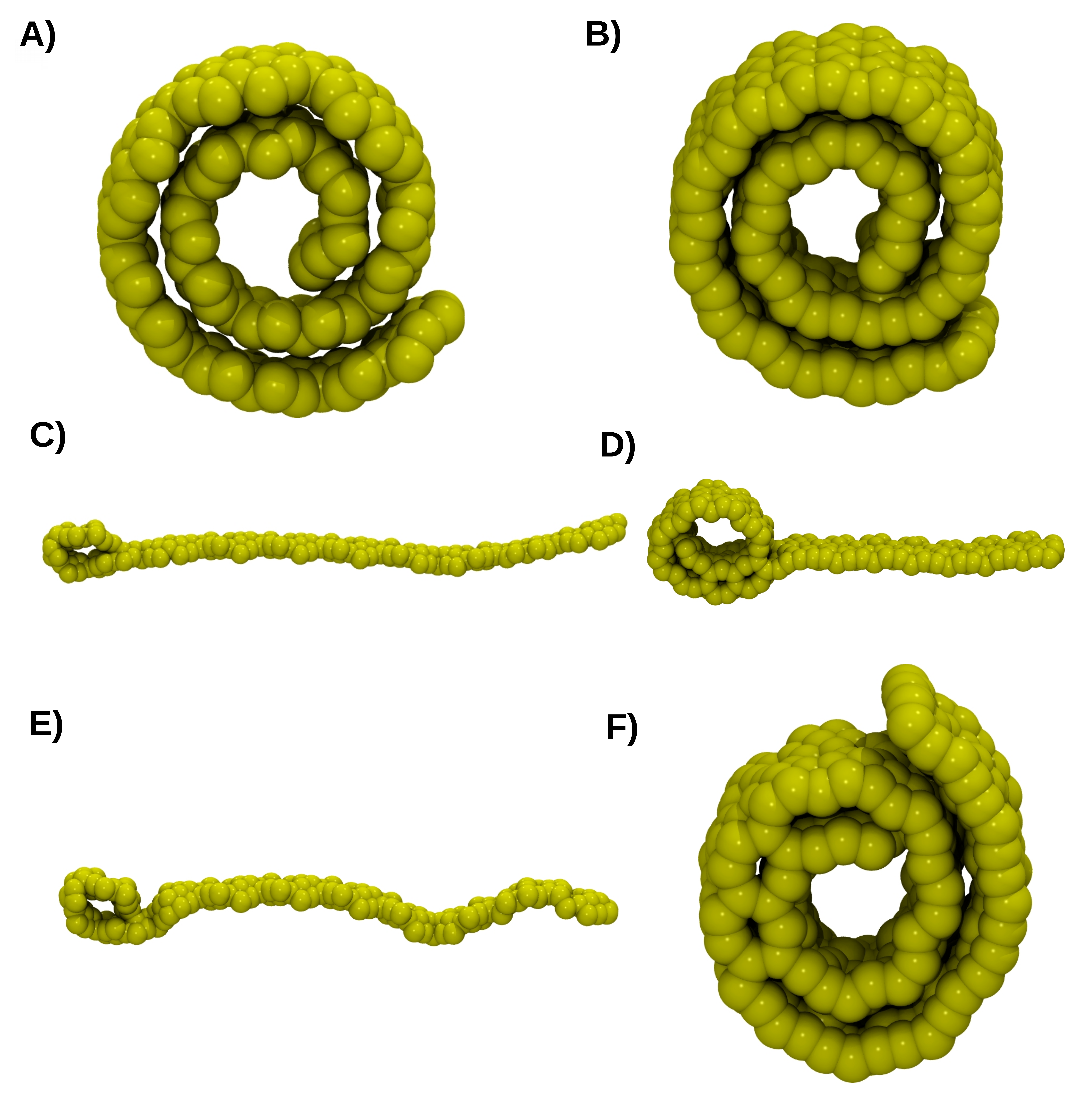}
\caption{ Representative MD snapshots showing the charge-induced unrolling process. (A) Initial configuration of an armchair nanoscroll before heating up to 600K. (C) unrolled armchair nanoscroll after being charged with 15 electrons at 600K. (E) Corrugated nanostrip after de-charging and cooling down the unrolled AC system. 
(B)Initial configuration of a zigzag nanoscroll before heating up to 600K. 
(D) Resulting configuration after a 15e charge at 600K for the zigzag model. 
(F) Nanoscroll after the rolling up triggered by de-charging and cooling down the unrolled zigzag system until 50K.  }
\label{fig:unroll_roll}
\end{figure}

In order to monitor the structural stability, MD simulations were performed considering various temperatures, up to 600K. Our results indicate that charge-neutral B-nanoscrolls are structurally stable for a large range of temperatures, presenting geometry modifications above room temperature. Such modifications do not compromise, however, the rolled-up characteristics.  
A relevant effect observed during the MD simulations above room temperatures, which is worth noting, is that a few interlayer bonds can be formed between boron atoms located at the center of two facing hexagons, as can be seen in Figure \ref{fig:interlayer-bonds}. These bonds have lengths of $\sim 1.84\AA$ for both chiralities considered in the present work and are able to preclude the unrolling process for low charge density and mild temperatures, namely, $T \lesssim 600K$. This behavior was not observed for nanoscrolls made of carbon (graphene, graphyne)\cite{Braga2004,Rurali_2006}. To further investigating the electroactuation effects on these systems, we carried MD simulations combining thermal energy injection, (used to increase the temperature), with charge injection. For neutral systems, the rolled configuration is stable up to 600K, but an unrolling effect can be obtained under a sufficiently large charge injection. For charge injections just above the threshold listed in Table \ref{table:energies}, the unrolling is blocked by interlayer bond formation, as cited above.
On the other hand, each chirality presents a threshold interval of injected charges ($15-17~e/cell$ for zigzag and $12-14~e/cell$ for the armchair case), above which the unrolling process is able to break those bonds and undergoes an almost complete unrolling, as shown in Figure \ref{fig:unroll_roll}. In that figure, there is a comparison between the unrolling process observed for an armchair nanoscroll and a zigzag B-nanoscroll. Both systems were heated up to 600K and maintained their rolled configurations for several ps, as shown in Figure \ref{fig:unroll_roll}(A) and \ref{fig:unroll_roll}(B). After 15e/unit cell injection, both systems unrolled, breaking almost all interlayer bonds, resulting in configurations of Figure \ref{fig:unroll_roll}(C) and \ref{fig:unroll_roll}(D). After charge removal and a cooling down to 50K, the armchair nanoscroll resulted in a corrugated configuration (Figure \ref{fig:unroll_roll}(E)), while the zigzag system returned to a completely rolled configuration (Figure \ref{fig:unroll_roll}(F)). Thus, the systems  exhibit remarkable reversible morphological changes, which can be controlled by charge injection/removal. The videos of the complete MD simulations, which illustrate the process of unrolling and subsequent rolling again for $2.5$ turns armchair and zigzag nanoscrolls, are included in the supplementary information.

\section{Conclusion}

In this work, we have proposed nanoscrolls based on $\alpha$-borophene nanosheets for the first time, and the calculations showed that they have structural and thermal stabilities. The armchair and zigzag conformations have similar electronic behavior due to their electronic bands dispersion and metallic behavior, however, they have slightly different interlayer distances. Zigzag conformation has small deformations in the innermost layer that point towards the outmost layer, resulting in atomic bonds due to the dangling bonds at the borders. 
It was also identified a giant electroactuation, which is more sensitive if compared to carbon nanoscrolls response to electrons injection, and they are entirely thermally reversible only for the zigzag conformation.

Since borophene sheets have already been obtained, in principle the same techniques used to produce carbon nanoscrolls could be used to obtain borophene nanoscrolls. We hope the present study will stimulate further studies along these lines.

{\color{blue} }

\begin{acknowledgement}

Guilherme S. L. Fabris thanks the postdoc scholarship financed by National Council for Scientific and Technological Development – CNPq grant \#150187/2023-8 and Fundação de Amparo à Pesquisa do Estado do Rio Grande do Sul – FAPERGS for the postdoc scholarship. Ricardo Paupitz acknowledges Fapesp for grants \#2018/03961-5 and \#2021/14977-2 and CNPq for grants \#437034/2018-6 and \#315008/2020-2. This work used resources of the "Centro Nacional de Processamento de Alto Desempenho em São Paulo (CENAPAD-SP)" and was also supported in part by resources supplied by the Center for Scientific Computing (NCC/GridUNESP) of São Paulo State University (UNESP). Douglas S. Galvão acknowledges CNPq for financial support. We thank the Center for Computing in Engineering and Sciences at Unicamp for financial support through the FAPESP/CEPID Grants \#2013/08293-7.
The authors acknowledge Professor Thomas Niehaus (Institut Lumière Matière, Claude Bernard University - Lyon 1) for having gently provided 
the wave function files used to obtain the orbital images shown throughout the text.
\end{acknowledgement}




\bibliography{bibliography}

\end{document}